\documentclass[12pt]{article}
\usepackage{amsmath}
\usepackage{amsfonts}
\usepackage{amssymb}
\usepackage{amsthm}
\usepackage[matrix,arrow,curve]{xy}

\newtheorem{theorem}{Theorem}[section]
\newtheorem{corollary}[theorem]{Corollary}

\hoffset= - 1.4in

\begin{document}
\title{
\begin{flushright}
\mbox{\normalsize \hfill ITEP/TH-03/13}
\end{flushright}
\vskip 20pt
Modified Hamilton formalism for fields
}
\author{ Danilenko Ivan\\ \\ITEP, Moscow, Russia }
\date{}
\maketitle

%\part*{ Modified Hamilton formalism for fields }

%\begin{flushright}
%Danilenko\\
%Ivan
%\end{flushright}
%\begin{center}
%October
%\end{center}

\abstract{ In Hamiltonian mechanics the equations of motion may be considered as a condition on the tangent vectors to the solution; they should be null-vectors of the symplictic structure. Usually the formalism for the field case is done by replacing the finite dimensional configuration space by an infinite dimensional one. In the present paper we work in worldsheet-targetspace formalism. The null-vectors of symplectic 2-form are replaced by null-polyvectors of a higher rank form on a finite dimensional manifold. The action in this case is an integral of a differential form over a surface in phase space. The method to obtain such a description from the Lagrange formalism generalizes the Legendre transformation. The requirement for this transformation to preserve the value of the action and its extremality leads to a natural definition of this procedure. }

\section{ Introduction }

\subsection{History of the problem}
%\subsubsection{ Hamiltonian mechanics }

It is widely known (see \cite{Arnold,ModGeom}), that given a configuration space $\mathcal{M}$, one can describe the Hamilton mechanics on its phase space $T^*(\mathcal{M})$ by a single function $H \colon T^*(\mathcal{M}) \to \mathbb{R}$. It is convenient to reformulate the well-known Hamilton equations of motion in terms of the natural symplectic structure $\omega = dp_i \wedge dq^i - dE \wedge dt $ (the function $-E$ is considered dual to $t$ like $p_i$ is dual for $q_i$; the sign "$-$" has historical reason) on the extended phase space $T^*(\mathcal{M}\times \mathbb{R})$. One can easily check, that the equation
\begin{equation}
i_{\dot{ \gamma } } \omega \vert_{E=H(p,q,t)} = 0,
\end{equation}
where $i$ means the inner product ($i_\xi \omega (\eta) := \omega(\xi,\eta)$) and $\dot{\gamma}$ is the tangent vector of a path, is equivalent to the standard set of Hamilton equations.

%\subsubsection{ The Hamiltonian formalism for fields }

The conventional Hamiltonian formalism of the field theory case arises from considering the space of all fields $\phi^i$ at a single moment as an infinite dimensional configuration space. The momentum $\pi^i$ is considered to be equal to the derivative of Lagrangian with respect to time derivatives of the fields: $\pi^i = \dfrac{\partial L}{\partial \dot{\phi}^i}$ The equations of motion can be written as a system of equations
\begin{equation}
\begin{split}
\dfrac{\partial \phi^i }{\partial t} &= \dfrac{\partial H }{\partial \pi^i } \\
\dfrac{\partial \pi^i }{\partial t} &= -\dfrac{\partial H }{\partial \phi^i }.
\end{split}
\end{equation}
or as a condition on time derivatives of any observable $F$:
\begin{equation}
\dot{F} = \lbrace F,H \rbrace.
\end{equation}

%\subsubsection{ The problems of the standard field formalism }

This conventional approach has an obvious defect. The time is considered as a coordinate, independent from the space ones. The description itself does not respect Lorentz invariance explicitly and it should be checked for the observables separately.

%\subsubsection{ The solution }

The problems associated with the separation of time can be cured by description considering all the spacetime coordinates as one worldsheet $\mathcal{T}$. In this case the states of the fields are maps from the worldsheet to the space $\mathcal{F}$ where the fields take their values. Such a map is considered as a surface in the targetspace $\mathcal{T\times F}$. The targetspace generalizes the conception of the extended configuration space. Thus this formalism replaces usual 1-dimensional paths of a particle by N-dimensional surfaces of the fields, there N is the dimension of spacetime. The symplectic structure is replaced by (N+1)-form $\omega = dp_{i_1,...,i_N}\wedge dq^{i_1} \wedge ... \wedge dq^{i_N}$, where $q^i$ are coordinates of the targetspace. The equations of motion are
\begin{equation}
i_\Xi \omega \vert_\Sigma = 0, 
\end{equation} 
where $\Xi$ is the tangent polyvector of a solution and $\Sigma$ is a certain submanifold in the phase space.

\subsection{The organization of the paper}

The present paper is organized as follows
\begin{itemize}
\item
In section \ref{sec:HMech} we describe our approach to Hamilton equations of motions. Section \ref{ssec:OMech} reformulates Hamiltonian mechanics in a specific form to make the following generalization as natural as it could be. Section \ref{ssec:FTMech} focuses on the modifications of the picture caused by transfer into the worldsheet case. A simple example of this mechanics is shown in section \ref{ssec:SF11} just to make it clear how it can actually work (it is a simplified version of the first example from section \ref{sec:examples} and the surface is not derived from the Lagrange function, but just guessed).
\item
Section \ref{sec:LT} describes the universal mechanism to obtain a Hamilton picture from a given Lagrange one. The ordinary Legendre transformation with modifications needed for the general case is presented. The description of new effects in the worldsheet case continues the section.
\item
Section \ref{sec:examples} contains 2 examples. The first one (section \ref{ssec:SF}) is the simpler one and describes the free scalar field. It contains the derivation of the surface guessed in section \ref{ssec:SF11} and the calculations for a scalar field in an n-dimensional spacetime. The second one (section \ref{ssec:ED}) is the case of electrodynamics in 1+1-dimension spacetime and it shows how a new effect -- Pl\"ucker relations -- modifies the formalism.
\end{itemize}

\section{ Hamilton mechanics in terms of degeneration }
\label{sec:HMech}
\subsection{Ordinary Hamilton mechanics}
\label{ssec:OMech}

Let us recall the description of Hamilton mechanics in terms of the symplectic structure (as described in \cite{Arnold}, also see \cite{ArnGiv}) and slightly modify it, considering energy as one of the coordinates. \\  
Geometric approach to Hamiltonian mechanics describes the motion in terms of the symplectic manifold $(M,\omega)$ and a function $H$ on it. The symplectic structure naturally yeilds a bijective mapping between vector and covector fields:
\begin{equation}
I\colon \xi \mapsto i_\xi \omega,
\end{equation}
where $i_\xi$ is a one-form given by $i_\xi \omega ( \eta ) := \omega ( \xi, \eta ) $ for any vector field $\eta$.\\
Such a duality maps 1-form $-dH$ to the corresponding Hamilton vector field $X_H$:
\begin{equation}
i_{X_H} + dH =0 \; \Leftrightarrow \; X_H = -I^{-1} dH \label{HamEq1}
\end{equation}
The equation (\ref{HamEq1}) is considered to be the equation of motion: the vector field $X_H$ describes a flux on $M$.\\
If the symplectic manifold is a cotangent bundle of some other manifold, the action form can be written as an integral of 1-form $p_idq^i - Hdt$ (in other cases this can be done locally due to $\omega$ being a closed form) and the equations of motions are derived by requiring the extremality of such an action. Equation (\ref{HamEq1}) is obtained as a consequence of this, with the help of Stokes' theorem.\\
$H$ is usually supposed to be a separate function and is not considered as a coordinate. However the approach treating $H$ and $p_i$ in similar way is very natural: the form of the action contains $p_i$ and $-H$ as objects of one nature. \\
For that purpose we consider Hamiltonian mechanics on the cotangent bundle of the extended configuration space $\mathcal{P}$, the product of the configuration manifold and $\mathbb{R}$ as time (actually, it is determined by a submanifold in $\mathcal{P}$ which is described further). The coordinate dual to time we denote by $-E$. Thus, the natural symplectic structure will be given by $\omega = dp_i\wedge dq^i - dE \wedge dt $.\\
Let us define a subsurface $\Sigma$ by the following equation
\begin{equation}
\Sigma \colon E - H(p,q,t) = 0.
\end{equation}
The restriction of $\omega$ to $\Sigma$ will lead to the degeneration of the symplectic structure. That degeneracy is related to the extremality of the action for the paths on $\Sigma$ by the Stokes' theorem, so the extremal paths must have velocity vectors on which $\omega$ vanishes. The null-vectors at non-singular point $x$ of $\Sigma$ form one-dimension linear subspace of $T_x \mathcal{P}$, determined by
\begin{equation}
i_\xi \omega \vert_\Sigma = 0, \label{HamEq2}
\end{equation}
where $\xi \in T_x \mathcal{P}$. All velocity vectors of a path with extremal action must belong to such subspaces. \\
%{\itshape The relation (\ref{HamEq2}) can be thought of as a duality of tangent hyperplanes, forming $\mathbb{P} ( T^*_x \mathcal{P} )$, since $\Sigma$ locally provides one, and tangent lines $\mathbb{P} ( T_x \mathcal{P} )$. This duality is a projectivization of familiar bijective mapping generated by $\omega$.} \\
In the present paper we work with Hamilton mechanics in this approach. It is provided by a restriction of non-degenerate symplectic form (denoted by $\omega$ above) to a submanifold (denoted by $\Sigma$ above) and the resulting degeneracy describes the motion in the sense of  (\ref{HamEq2}). The modification for the field formalism requires redefinition of $\omega$ and the tangent vector, as well as an algorithm of derivation $\Sigma$ from the Lagrangian density. The first modification is done in section (\ref{ssec:FTMech}) and the second one is described in section (\ref{sec:LT}).

\subsection{Hamilton mechanics in field theory}
\label{ssec:FTMech}

In this section we are trying to obtain a description of motion analogous to the one described in the previous section, this time in the case of field theory. Our approach is based on considering the field theory in worldsheet-targetspace formalism. This approach is also appropriate for objects like strings. The only difference is that worldsheet would have a non-trivial topology by itself, but in our consideration that effects in the same way as a topology of configuration space. \\
For that aim we must modify the objects we used in mechanics to obtain an applicable picture. Let us denote a worldsheet (an analogue of time) as $\mathcal{T}$ and a space where fields take their values (an analogue of configuration space) as $\mathcal{F}$. The variation problem is formulated for mappings of $\mathcal{T}$ to $\mathcal{F}$. They can be considered as surfaces of dimension $N = dim \mathcal{T}$ in the targetspace $\mathcal{T \times F}$. The new features are
\begin{enumerate}
\item
 The local properties of path used to be described by a velocity vector. It should be replaced by a tangent polyvector from $\Lambda^N T \left( \mathcal{T \times F} \right)$.
\item
 The action was obtained by integration of the natural 1-form along the path in the phase space. To be integrated, now it should be $N$-form and $\Lambda^N T^* \left( \mathcal{T \times F} \right)$ is naturally equipped with such a form in the same way as $ T^* \left( \mathcal{T \times F} \right)$ is equipped with the natural one-form (see~\cite{Arnold}). So the "phase space" should be $\Lambda^N T^* \left( \mathcal{T \times F} \right)$. Any element of this is a differential form
\begin{equation}
\sum\limits_{I \in A} p_I \left( \bigwedge\limits_{i \in I}dx^i \right)
\end{equation}
where $x^i$ are coordinates on $ \mathcal{T \times F} $, A is a set of multiindices of length $N$, so $x^i$ and $p_I$ are coordinates on $\Lambda^{N} T^*\left( \mathcal{T \times F} \right)$.
\item
The action for an $N$-dimension surface $\Gamma$ in the phase space is given by the integral
\begin{equation}
S = \int\limits_\Gamma \alpha,
\end{equation}
where
\begin{equation}
\alpha = \sum\limits_{I \in A} p_I \left( \bigwedge\limits_{i \in I}dx^i \right),
\end{equation}
This form can be defined in a coordinate-independent way: at a point $P \in \Lambda^{N} T^* \left( \mathcal{T \times F} \right)$, the form $\alpha$ is defined by $\alpha(\xi) = P(\pi_* \xi)$ (~$\pi $~is~the~natural mapping $\pi\colon \Lambda^{N} T^* \left( \mathcal{T \times F} \right) \to \mathcal{T \times F}$ ).\\
The symplectic form is replaced by a $(N+1)$-form $\omega$, an exterior derivative of the action form. It is given by
\begin{equation}
\omega := d\alpha = \sum\limits_{I \in A} dp_I\wedge \left( \bigwedge\limits_{i \in I}dx^i \right).
\end{equation}
For a fixed surface $\Sigma$ in the phase space, $\omega$ should vanish on extremal subsurfaces of $\Sigma$ (it can be derived by the Stokes' theorem): a subsurface of $\Sigma$ is extremal w.r.t. the action, iff for all its tangent polyvectors $ \Xi $ the equation
\begin{equation}
i_\Xi \omega \vert_\Sigma = 0
\end{equation}
is satisfied. Thus surface $\Sigma$ defines the motion in the same way as in (\ref{HamEq2}).
\item
The last special feature for the worldsheet case is that all tangent polyvectors should be decomposable. The polyvector is decomposable, iff it can be presented as an exterior product of vectors.
\end{enumerate}

\subsection{Example}
\label{ssec:SF11}

Now we will show that this approach works in a simple case. Here in a special case we are going to guess a surface $\Sigma$ which provides the proper equations of motion.\\
Let us consider a free scalar field in 1+1-dimensional spacetime $ \mathcal{T} $. In this case $\mathcal{F} $ is one-dimensional (the field is scalar) and we denote the corresponding coordinate as $\phi$. Two space-time coordinates are $x^0, \; x^1$. The phase space is $\Lambda^2 T^*\left( \mathbb{R}^3 \right)$. We should obtain Klein-Gordon equation $\bigtriangleup \phi = 0$, from the null-polyvectors of $\omega$ after restriction to a subsurface $\Sigma$. In this particular case
\begin{equation}
\omega = dp_{01} \wedge dx^0 \wedge dx^1 + dp_{\phi 0}\wedge d\phi \wedge dx^0 + dp_{\phi 1}\wedge d\phi \wedge dx^1.
\end{equation}
(in expressions $p_{\phi 0}$ and $p_{\phi 0}$ index $\phi$ corresponds to the coordinate $\phi$ of $\mathcal{F} $)\\
Any polyvector in $\Lambda^{ dim \mathcal{T} } T \left( \Lambda^{ dim \mathcal{T} } T^* \left( \mathcal{T \times F} \right) \right) $ ("tangent bundle of phase space") with non-degenerate projection on $\mathcal{T} $ can be written in the following form
\begin{equation}
\begin{split}
\Xi = & C \left( \dfrac{\partial\:}{\partial x^0} + f_0 \dfrac{\partial\:}{\partial \phi} + \pi_{01;0} \dfrac{\partial\:}{\partial p_{01}} + \pi_{\phi 0; 0} \dfrac{\partial\:}{\partial p_{\phi 0}} + \pi_{\phi 1; 0} \dfrac{\partial\:}{\partial p_{\phi 1}} \right) \wedge \\ 
& \wedge \left( \dfrac{\partial\:}{\partial x^1} + f_1 \dfrac{\partial\:}{\partial \phi} + \pi_{01;1} \dfrac{\partial\:}{\partial p_{01}} + \pi_{\phi 0; 1} \dfrac{\partial\:}{\partial p_{\phi 0}} + \pi_{\phi 1; 1} \dfrac{\partial\:}{\partial p_{\phi 1}} \right)
\end{split}
\end{equation}
For such a polyvector $n$-form $i_\Xi \omega$ is given by
\begin{equation}
\begin{split}
i_\Xi \omega = & C ( dp_{01} - f_1 dp_{\phi 0} + f_0 dp_{\phi 1} + (\pi_{\phi 0; 1} - \pi_{\phi 1; 0})d\phi + \\ &+ (f_1 \pi_{\phi 0;0} - f_0 \pi_{\phi 0;1} - \pi_{01; 0} )dx^0 +  \\  &+ (f_1 \pi_{\phi 1;0} - f_0 \pi_{\phi 1;1} - \pi_{01; 1} ) dx^1 ).
\end{split}
\end{equation}
Suppose $\Sigma$ is defined by one equation 
\begin{equation}
\Sigma: F=0.
\end{equation}
In this case $i_\Xi \omega \vert_\Sigma $ vanishes iff $i_\Xi \omega$ is proportional to $dF$:
\begin{equation}
 i_\Xi \omega = \alpha dF.
\end{equation}
Expressing $p_{01}$ locally from the implicit function $ F = 0 $
\begin{equation}
p_{01} = \Phi(p_{\phi 0}, p_{\phi 1}, \phi )
\end{equation}
(the dependence on $x^0$ and $x^1$ may be added, but we don't do it as Klein-Gordon equation respects shifts) one obtains the requirement of degeneracy:
\begin{equation}
 i_\Xi \omega = \alpha ( dp_{01} - d\Phi ).
\end{equation}
Using $ f_i = \dfrac{\partial \phi}{\partial x^i}, \; \pi_{I,j} = \dfrac{\partial p_I }{\partial x^j} $ that leads to
\begin{align}
-\dfrac{\partial \phi}{\partial x^1} &=-\dfrac{\partial \Phi}{\partial p_{\phi 0}} \label{Part2_1}\\
\dfrac{\partial \phi}{\partial x^0} &=-\dfrac{\partial \Phi}{\partial p_{\phi 1}} \label{Part2_2} \\
\dfrac{\partial p_{\phi 0} }{\partial x^1} - \dfrac{\partial p_{\phi 1} }{\partial x^0} &=-\dfrac{\partial \Phi}{\partial \phi} \label{Part2_3} \\
\dfrac{\partial p_{\phi 0} }{\partial x^0}\dfrac{\partial \phi}{\partial x^1} - \dfrac{\partial p_{\phi 0} }{\partial x^1}\dfrac{\partial \phi}{\partial x^0} - \dfrac{\partial \Phi}{\partial \phi}\dfrac{\partial \phi}{\partial x^0} - \dfrac{\partial \Phi}{\partial p_{\phi 0}}\dfrac{\partial p_{\phi 0} }{\partial x^0} - \dfrac{\partial \Phi}{\partial p_{\phi 1}}\dfrac{\partial p_{\phi 1} }{\partial x^0} &=-\dfrac{\partial \Phi}{\partial \phi}\dfrac{\partial \phi}{\partial x^0} \label{Part2_4} \\
\dfrac{\partial p_{\phi 1} }{\partial x^0}\dfrac{\partial \phi}{\partial x^1} - \dfrac{\partial p_{\phi 1} }{\partial x^1}\dfrac{\partial \phi}{\partial x^0} - \dfrac{\partial \Phi}{\partial \phi}\dfrac{\partial \phi}{\partial x^1} - \dfrac{\partial \Phi}{\partial p_{\phi 0}}\dfrac{\partial p_{\phi 0} }{\partial x^1} - \dfrac{\partial \Phi}{\partial p_{\phi 1}}\dfrac{\partial p_{\phi 1} }{\partial x^1} &=-\dfrac{\partial \Phi}{\partial \phi}\dfrac{\partial \phi}{\partial x^1} \label{Part2_5}
\end{align}
The last two equations are satisfied if the first two ones hold.\\
Finally, the required surface can be guessed since we have not yet provided a technique to calculate it\footnote{The derivation of the surface equation in section \ref{ssec:SF} provides the same result by an algorithm presented in section \ref{sec:LT}. }. The equations (\ref{Part2_1}) - (\ref{Part2_3}) with
\begin{equation}
\Phi = \frac{1}{2} p^2_{\phi 0} - \frac{1}{2} p^2_{\phi 1}
\end{equation}
yield a system
\begin{align}
-\dfrac{\partial \phi}{\partial x^1} &= - p_{\phi 0} \\
\dfrac{\partial \phi}{\partial x^0} &= p_{\phi 1} \\
\dfrac{\partial p_{\phi 0} }{\partial x^1} - \dfrac{\partial p_{\phi 1} }{\partial x^0} &= 0.
\end{align}
which have the Klein-Gordon equation for metric $(dx^0)^2-(dx^1)^2$ as a corollary:
\begin{equation}
\left( \dfrac{\partial \; }{\partial x^1 } \right) ^2 \phi - \left( \dfrac{\partial \; }{\partial x^0 } \right) ^2 \phi = 0.
\end{equation}
\emph{Remark.} Equations (\ref{Part2_4}) and (\ref{Part2_5}) are equivalent to (\ref{Part2_3}), if (\ref{Part2_1}) and (\ref{Part2_2}) are true. There is an analogous situation in the general case (see section \ref{ssec:SF}).\\

In this section the subsurface was just guessed. In the following part we are going to show how to obtain it from the Lagrange formalism. It is an analogue of the Legendre transformation.

\section{The Legendre transformation}
\label{sec:LT}

In this section a modification of Legendre transformation is presented. It constructs a surface $\Sigma$ in the phase space for a given Lagrange function. It works in the same way as the conventional one in the cases where the conventional one can be applied. It also provides a technique which was essential is \ref{ssec:SF11} to derive the surface from the Lagrange function.

\subsection{Worldline case}
\subsubsection{Function $\Lambda$}

Action is defined by an integral over a path. A natural requirement for it is that it mustn't depend on parametrization. Nevertheless usually only the parametrization by time is used. To correct this we should first think which type of objects can be integrated over 1-dimensional manifold. First, it should be a function on a tangent space (as 1-form is). Second, it should be a homogeneous function of degree 1 on each tangent space. Usually it is supposed to be a 1-form, but it is an unnecessary restriction (the integration doesn't need to respect the addition of vectors). So action can be obtained as an integral of a homogeneous function of degree 1 over a path. Denote that function as $\Lambda$. \\
Traditionally the function for integration is written in form $L dt$. We should see the relation between $L$ and $\Lambda$. \\
Suppose a timeline $\mathcal{T}$ and a manifold $M$ with a Lagrange function $L$ on it is given. $L$ can be regarded as a function $L^*$, acting on a subspace of $T(\mathcal{T} \times M  )$ defined by $dt = 1$ in the following way:
\begin{equation}
L^* \left( q^i, t, \dfrac{\partial\:}{\partial t} + \chi^i \dfrac{\partial\:}{\partial q^i} \right) = L \left( q^i, t, \chi^i \right) .
\end{equation}
$L^*$ can be extended to a function $\Lambda \colon T(\mathcal{T} \times M  ) \to \mathbb{R}$, homogeneous on each tangent space. The corresponding formula is the following one (we omit the dependence on $q_i$ and $t$ in the parameters like it is usually done for differential forms):
\begin{equation}
\Lambda \left( \tau \dfrac{\partial\:}{\partial t} + \chi^i \dfrac{\partial\:}{\partial q^i} \right) = \tau L^* \left( \dfrac{\partial\:}{\partial t} + \dfrac{\chi^i}{\tau} \dfrac{\partial\:}{\partial q^i} \right) .
\end{equation}
for any vector $\tau \dfrac{\partial\:}{\partial t} + \chi^i \dfrac{\partial\:}{\partial q^i} \in T(\mathcal{T} \times M  )$.
%\emph{Remark.} {\itshape This formula can be interpreted as $\Lambda \sim L \cdot dt$, because the value of $dt$ on tangent vector is $\tau$. This is obvious, since integral of both must be the action.}\footnote{We use "$\sim $" instead of  "$=$" as far as $L$ and $\Lambda $ have different arguments and can't be equal as functions. This also caused extra steps and to avoid such formal difficulty. The reader should not be confused by that. }\\

So the action on a path $\gamma(\tau )$ ($\tau$ is any parameter of that path) is the integral
\begin{equation}
S = \int\limits_\gamma \Lambda := \int d\tau \cdot \Lambda\left( \frac{d\gamma}{d\tau} \right) .
\end{equation}
The homogeneity of $\Lambda $ causes independence of $S$ from the choice of parameter $\tau$.\\
Usually reparametrization invariance of action is considered as a special case. Here we show that any Lagrange function has a corresponding reparametrization invariant $\Lambda $.

\subsubsection{Transformation}

Now we will focus only on one tangent space $V := T_{t,q}(\mathcal{T} \times M )$. What properties do we expect from the Legendre transformation $ \pi \colon V \to V^* $? There are two requirements:
\begin{itemize}
\item
First, if path $\gamma (\tau )$ has action $ S = \int \limits_\gamma \Lambda $, its image in the phase space  $ \tilde{ \gamma } (\tau ) := ( \gamma (\tau ), \pi ( \dot \gamma (\tau ) ) )  $ should have the same action $ S = \int \limits_{\tilde \gamma } pdq $ (time is also included in $ pdq $). So
\begin{equation}
\forall \xi \in V \quad \langle \pi(\xi), \xi \rangle = \Lambda(\xi), \label{Crit1}
\end{equation}
where $\langle,\rangle $ is the coupling of a covector and a vector.
\item
The second restriction is that the function $f(\eta) = \langle \pi (\eta), \xi \rangle$ must have an extremal value at point $ \eta = \xi $, so
\begin{equation}
 \forall \eta \in V \quad \langle d\pi |_\xi(\eta), \xi \rangle = 0. \label{Crit2}
\end{equation}
This requirement must be added due to the appearence of the extra degrees of freedom. For a curve in the
configuration space (which is the object under variation in Lagrange mechanics) we could not vary velocities independently. However for a curve in the phase
space one can change its momentums and coordinates without any correlation. The
presented rule is equivalent to extremality of the action for the path 
$ \tilde{ \gamma } (\tau ) := ( \gamma (\tau ), \pi ( \dot \gamma (\tau ) ) )  $ 
in the set of all paths in the phase space having projection $\gamma (\tau )$ in
the configuration space.
\end{itemize}

If these conditions hold, one can easily prove\\
\begin{theorem}
If $\gamma (\tau )$ corresponds to an extremal value of the action in Lagrange mechanics, then $ \tilde{ \gamma } (\tau ) := ( \gamma (\tau ), \pi ( \dot \gamma (\tau ) ) )  $ corresponds to an extremal value of the action $ S = \int \limits_{\tilde \gamma } pdq $, i.e. the map $ \gamma (\tau ) \to \tilde{ \gamma } (\tau ) $ sends extremal paths to extremal ones.
\end{theorem}

\begin{corollary}
If path $\gamma $ satisfies the Lagrange-Euler equations, it satisfies the equation (\ref{HamEq2}) where for the surface $\Sigma$ one takes the image of $\pi$.
\label{Corollary}
\end{corollary}

 Since (\ref{HamEq2}) is equivalent to the extremality of the action for all paths in $\Sigma $ one may consider the image of $\pi$ as $\Sigma $ because of Corollary \ref{Corollary}. $\Sigma $ has a description following by the properties of $\pi$. We will describe it in terms of dual algebraic varieties (for the definition see \cite{Harris}). \\

The two presented conditions on $\pi$ can be expressed as one requirement: $\pi (\xi)$ should be a double zero of the function (defined on $\Sigma $)
\begin{equation}
 F_\xi(p) = \langle \pi,\xi \rangle - \Lambda(\xi). \label{Fdef}
\end{equation}
In particular, $F_\xi(p) = 0$ is (\ref{Crit1}) and $dF\vert_\xi(p) = 0$ is  equivalent to (\ref{Crit2}).\\
 
\emph{Remark}. $\Sigma$ and the presented requirement are enough to describe $\pi (\xi)$. In some cases in place of $\pi (\xi)$ even something multivalued can arise (an image of a single point can even be a manifold). Due to that requirement (\ref{Fdef}) actually gives more freedom. This situation appears, for example, in section \ref{ssec:ED}.  \\

The function (\ref{Fdef}) can be made bihomogeneous by adding a new variable $\Pi$: 
\begin{equation}
\tilde{F}(P,X) = P(X),
\end{equation} 
where $P$ is a covector $ ( \Pi,p ) \in \mathbb{R} \oplus V^*$ and $X $ is a vector $ ( \Lambda, \xi ) \in \mathbb{R} \oplus V$. $\tilde{F}$ is just a pairing of a vector with a covector. The relation of $F$ and $\tilde{F}$ is the following one
\begin{equation}
F_\xi(p) = \tilde{F} \left. \left( {\Pi \choose p} , { \Lambda \choose \xi } \right) \right| _{\Pi = -1,\: p \in \Sigma ,\: \Lambda = \Lambda(\xi) } .
\end{equation} 
As one can see the condition $\tilde{F}(P,X_0) = 0$ defines a hyperplane $H$ in $\mathbb{P} \left( \mathbb{R} \oplus V^* \right)$, associated with an equivalence class vector $[X_0]$ (it is a covector for $\left( \mathbb{R} \oplus V^* \right) $, so it is defined by it zeroes up to a multiplication by nonzero scalar). Let us define the projectivization of $\Sigma \subset V^* $ as $\Upsilon \subset \mathbb{P} \left( \mathbb{R} \oplus V^* \right)$ ( $\Sigma$ is supposed to be $\Upsilon $ in an affine map $\Pi = -1$ ). Then $\tilde{F}(P,X_0)$ being restricted to $\Upsilon $ has a double zero at $[P_0] \in \Upsilon $ ( and, equivalently, has a double zero at the point of $\Sigma$, corresponding to $[P_0]$ ), iff $H$ is tangent to $\Upsilon $ at $[P_0]$. Thus the vector $\xi\in V$ is mapped to $p\in V^*$, iff $[X_0] = [ { \Lambda (\xi) \choose \xi } ]$ corresponds to a hyperplane tangent to $\Upsilon $ at $[P_0] = [ { -1 \choose p } ]$. Denote the set of all $[X_0]$ corresponding to tangent  hyperspaces of $\Upsilon $ as $\Upsilon^*$. $\Upsilon^*$ is called dual manifold to $\Upsilon$. This correspondence is an involution: the manifold and its double dual are equal. \\
Let us denote as $\Xi $ all the points of $\mathbb{P} \left( \mathbb{R} \oplus V \right)$ which have the form $[ { \Lambda (\xi) \choose \xi } ]$ (this is the projection of the graph of $\Lambda$ to $\mathbb{P} \left( \mathbb{R} \oplus V \right)$). Since $\Upsilon$ is formed by all the images of $\xi\in V$, $\Xi \subset \Upsilon^* $. So $\Upsilon = (\Upsilon^*)^* \subset \Xi^* $.\\
Any subsurface $\Upsilon \subseteq \Xi^* $ which has for each point $[X_0] \in \Xi $ at least one corresponding point $[P_0] \in \Upsilon$ can be chosen to obtain proper equations of motion. The certain choice of $[P_0] \in \Upsilon$ may be considered as a choice of gauge.\\
In any situation $\Upsilon = \Xi^* $ may be chosen. It is the simplest case, but may lead to multivalued mapping $\pi$. If one takes $\Upsilon \subset \Xi^* $, the number of values decreases. As shown later, in the case of 1d-time and convex Lagrange function the choice $\Upsilon = \Xi^* $ leads to a single-valued function $\pi$, so it is the only choice possible since the number of values can't be decreased. Further on we consider that choice.\footnote{Actually $\Xi^*$ is a well-defined algebraic variety if $\Xi$ is an algebraic variety, but $\Xi^*$ may fail to be a smooth manifold even if $\Xi$ is a smooth manifold. However for a wide class of Lagrange functions either $\Xi$ is an algebraic variety (if Lagrange function is a ratio of polynomials and in some other cases) or $\Xi^*$ is a manifold. Further we consider that $\Xi^*$ is well-defined. (see \cite{Harris}) }\\

Summarizing all mentioned above, we visualize the procedure by the following diagram (with  $\Upsilon = \Xi^* $)
%One can prove, that $F_\xi$ has a double zero at $p$ iff a hyperplane in $\mathbb{P} \left( \mathbb{R} \oplus V \right)$, associated with a covector $(-1,p)$, is tangent to a graph of function $\Lambda$ at $(\Lambda, \xi)$. Because of homogeneity of $\Lambda$ its graph can be projected to a manifold in $\mathbb{P} \left( \mathbb{R} \oplus V \right)$. Let us denote this submanifold as $\Xi$. The set of all tangent hyperspaces in $\mathbb{P} \left( \mathbb{R} \oplus V^* \right)$ forms a manifold $\Xi^*$ called dual to $\Xi$ \footnote{Actually $\Xi^*$ is a well-defined variety if $\Xi$ is a variety, but $\Xi^*$ may fail to be a manifold even if $\Xi$ is a manifold. However for a wide class of Lagrange functions either $\Xi$ is a variety ( if Lagrange function is a ratio of polynomials and in some other cases ) or $\Xi^*$ is a manifold. Further we consider that $\Xi^*$ is well-defined. } ( see \cite{Harris} ). The usual phase space can be found in affine map $\Pi = -1$. So, we obtained a formulation of Legendre transformation in terms of dual manifolds.

%This procedure can be visualized by the following diagram
\begin{equation}
\xymatrix{
\Xi \subset \mathbb{P} \left( \mathbb{R} \oplus V \right) \ar@{<->}[rr]^{(c)} && \Xi^* \subset \mathbb{P} \left( \mathbb{R} \oplus V^* \right) \ar@{->}[d]^{(d)} \\
\Lambda \colon V \to \mathbb{R} \ar@{->}[u]^{(b)} && \Sigma \subset  \mathbb{A} \left( V^* \right) \ar@{->}[d]^{(e)}\\
Lagrange\: function \ar@{<-->}[rr]^{Legendre}_{transformation} \ar@{->}[u]^{(a)} && Hamilton\: function
} \label{Diagram}
\end{equation}
The dashed line shows the duality provided by the Legendre transformation. In our approach we have the duality between $\Xi$ and $\Xi^*$ ((c) is an involution). The steps which connect Lagrange and Hamilton pictures are
{\renewcommand{\theenumi}{\alph{enumi}}
\renewcommand{\labelenumi}{(\theenumi)}
\begin{enumerate}
\item
Construction of homogeneous function $\Lambda$ from the given Lagrange function. The existence of that function is just based on the independence of the action value from the parametrization of the path - it depends only from its form. The classical parametrization by the time is not the unique chose.
\item
Note that the graph of $\Lambda$ (which lies in $\mathbb{R} \oplus V$) is mapped to itself by any dilatation of the vector space due to the homogeneity of order one of $\Lambda$: if ${ \Lambda \choose \xi }$ is a point of graph, ${ \alpha \Lambda \choose \alpha \xi }$ is also a point of graph. So that graph can be projected to a submanifold $\Xi \subset \mathbb{P} \left( \mathbb{R} \oplus V \right) $ by a natural projection $\mathbb{R} \oplus V \setminus {0} \to \mathbb{P} \left( \mathbb{R} \oplus V \right). $
\item
A standard duality of projective varieties can be performed on $\Xi$ to obtain $\Xi^* .$
\item
Setting a coordinate $\Pi$, dual to $\Lambda$, equal to $-1$, one obtains an affine map $\mathbb{A} \left( V^* \right)$ of projective space $ \mathbb{P} \left( \mathbb{R} \oplus V^* \right) $ with a submanifold $\Sigma$ as the image of $\Xi^* .$
\item
If $\Sigma$ has codimension 1 (that corresponds to non-degenerate Lagrange functions), using the implicit function theorem, one can locally express one coordinate as a functions of the others. Usually it is given in form $E = H(p_i,q^i,t)$ and that defines function $H$. 
\end{enumerate}
}

\emph{Remark.} Suppose a Lagrange function $\mathcal{L} $ which is convex as a function of velocities is given. We are going to show that in this case the presented algorithm acts as a traditional Legendre transformation. The steps are the same as in diagram (\ref{Diagram}). We omit the dependence of functions on coordinates and time and focus only on dependence on velocities.

{\renewcommand{\theenumi}{\alph{enumi}}
\renewcommand{\labelenumi}{(\theenumi)}
\begin{enumerate}
\item
In this case function $\Lambda $ is

\begin{equation}
\Lambda \left( \tau \dfrac{\partial\:}{\partial t} + \chi^i \dfrac{\partial\:}{\partial q^i} \right) = \tau L \left( \dfrac{\chi^i}{\tau} \right)
\end{equation}
as it was previously derived.
\item
One can check the homogeneity of $\Lambda $:
\begin{equation}
\Lambda \left( \alpha \left[ \tau \dfrac{\partial\:}{\partial t} + \chi^i \dfrac{\partial\:}{\partial q^i} \right] \right) = \alpha \tau L \left( \dfrac{\alpha \chi^i}{\alpha \tau} \right) = \alpha \Lambda \left( \tau \dfrac{\partial\:}{\partial t} + \chi^i \dfrac{\partial\:}{\partial q^i} \right)
\end{equation}
and this proves that the graph of $\Lambda $ defines a submanifold $\Xi $ of the projective space with homogeneous coordinates $[\Lambda : \tau : \chi^1 : ... : \chi^n ]$.
\item
A point of the dual projective space with homogeneous coordinates $[\Pi : -E : p_1 : ... : p_n ] $ defines a hyperplane tangent to a point of $\Xi $, iff $(\Pi , -E , p_1 , ... , p_n )$ is proportional to covector 
\begin{equation}
d\left( \Lambda - \tau L \left( \dfrac{\chi^i}{\tau} \right) \right) = \left( 1 , - L \left( \dfrac{\chi^i}{\tau} \right) + \dfrac{\chi^j}{\tau} \partial_j L \left( \dfrac{\chi^i}{\tau} \right) , - \partial_1 L \left( \dfrac{\chi^i}{\tau} \right) , ... , - \partial_n L \left( \dfrac{\chi^i}{\tau} \right) \right)
\end{equation}
with any $\tau , \chi^i $. This is equivalent to
\begin{equation}
rank
\begin{pmatrix}
1 & - L \left( \dfrac{\chi^i}{\tau} \right) + \dfrac{\chi^j}{\tau} \partial_j L \left( \dfrac{\chi^i}{\tau} \right) & - \partial_1 L \left( \dfrac{\chi^i}{\tau} \right) & \cdots & - \partial_n L \left( \dfrac{\chi^i}{\tau} \right) \\
\Pi & -E & p_1 & \cdots & p_n
\end{pmatrix} = 1.
\end{equation}
All the $2\times 2$ minors must vanish: 
\begin{align}
0 & = 
\begin{vmatrix}
1 & - L \left( \dfrac{\chi^i}{\tau} \right) + \dfrac{\chi^j}{\tau} \partial_j L \left( \dfrac{\chi^i}{\tau} \right) \\
\Pi & -E
\end{vmatrix}
 = - E + \Pi \left[ L \left( \dfrac{\chi^i}{\tau} \right) - \dfrac{\chi^j}{\tau} \partial_j L \left( \dfrac{\chi^i}{\tau} \right) \right] , \\
 0 & = 
\begin{vmatrix}
1 & - \partial_j L \left( \dfrac{\chi^i}{\tau} \right) \\
\Pi & p_j
\end{vmatrix}
 = p_j + \Pi \partial_j L \left( \dfrac{\chi^i}{\tau} \right) .
\end{align}
In the case of convex $L$ the mapping  $[\Lambda :\tau :\chi^i] \to [\Pi : -E: p_i ] $ is single-valued as it was said previously.\\
Denoting $\dfrac{\chi^i}{\tau} $  as $v^i$, we obtain a parametric description of $\Xi^* $ by a system:
\begin{gather}
- E + \Pi \left[ L \left( v^i \right) - v^j \partial_j L \left( v^i \right) \right] = 0 , \\
p_j + \Pi \partial_j L \left( v^i \right) = 0
\end{gather}
(parametrized by $v^i$).
\item
In this step we obtain a subsurface $\Sigma$ in cotangent space with coordinates $(-E,p_1,...,p_n)$.\\
%As described, it is situated at the affine map $\Pi = -1 $.
$\Sigma$ is the subset of $\Xi^* $ covered by the affine chart $\Pi = -1 $. So $\Sigma$ is parametrized by $v^i$ as follows
\begin{gather}
- E = \left[ L \left( v^i \right) - v^j \partial_j L \left( v^i \right) \right] , \\
p_j = \partial_j L \left( v^i \right)
\end{gather}
One can note that $ L \left( v^i \right) - v^j \partial_j L \left( v^i \right) $ is equal to traditional $-H(p_i)$ and this system provides Legendre transformation.\\
$\Sigma$ is defined by equation
\begin{equation}
E - H(p_i) = 0. \label{Sigma1d}
\end{equation}
\item
The Hamilton function is obtained from $\Sigma$ if $E$ is explicitly expressed from $p_i$ as it is in (\ref{Sigma1d}). So $H(p_i)$ in this equation is the Hamilton function.
\end{enumerate}
}

\subsection{Advantages of the new description}

This description has extra possibilities. They are tightly connected with the appearance of the varieties on the both sides of arrow (c). The classic approach needs both varieties to be of codimension 1. Both $\Xi $ and $\Xi^* $ may have a higher codimension in several situations. Here we list some of them
\begin{itemize}
\item
\emph{ $\Xi$ has higher codimension}. In this case extra equations on $\Xi $ appear. Usually they appear if Lagrange function due to some reason is defined on submanifold defined by equations $f_\alpha (q,v) = 0$ (we suppose that $f_\alpha (q,\lambda v) = 0$ if $f_\alpha (q,v) = 0$ to enable projection to the projective space). In our work they appear in the case of fields when not all the tangent polyvectors are decomposable.\\
In case of Lagrange function with higher derivatives it may be considered as a function on jet bundle which is s subbundle of $T^k(\mathcal{T\times M})$. But this subbundle isn't defined by homogeneous functions: in case of second derivatives $L$ is defined for $(x,v,\dot{x},\dot{v})$ iff for $\dot{x} = v$. That defines a plane in vector space (this plane depends on coordinate $v$). It is interesting if a modification to cope with such nonhomogeneous situations exists (for existing Hamiltonian formalisms in these cases see, for example \cite{Morozov})
\item
In cases with non-convex Lagrange function $\Xi^* $ may have higher codimension (in this situation $\Xi$ is called {\slshape defected}, see \cite{Harris}). This effect in mechanics is called Dirac constraints (see \cite{Tyutin}). Despite the system is not defined by a single function such as Hamilton function, the formalism with equation (\ref{HamEq2}) still works. The modified Legendre transformation properly describes the system.\\
The non-convexity of  Lagrange function $L$ may be caused by reparametrization invariance of $L$. As one can check this leads to independence of $\Lambda$ from formal time, i.e. $\Xi$ is a cylinder. Thus $\Xi^* $ lies is a hypersurface and this is a Dirac constraint (compare with \cite{DBS} in case of first derivatives).
\end{itemize}

Now we are going to revisit the choice $\Upsilon = \Xi^* $ made in the previous section. In a theory with $N$ constraints $f_\alpha (q,v) = 0$ an arbitrary point has an $N$-dimentional linear subspace of corresponding points in $\Xi^* $. So by choosing a specific $\Upsilon \subset \Xi^* $ one may make that correspondence to be a bijective mapping. Nevertheless the choice $\Upsilon = \Xi^* $ is still the most convenient due to the following reason. The paths in the phase space need to have a velocity vector which has a projection from the variety $f_\alpha (q,v) = 0$. In the case of arbitrary $\Upsilon $ this restriction should be checked explicitly. However in case $\Upsilon = \Xi^* $ these equations are consequences of path extremality: if an extremal path in the set of all paths with momenta from $\Upsilon $ has momentum $\pi $, its velocity $v$ is from $\Upsilon^* $ since the action is their pairing $\pi(v)$. If one chooses $\Upsilon = \Xi^* $, $v$ belongs to $\Upsilon^* = \Xi $ and $f_\alpha (q,v) = 0$ are satisfied. Further we again consider that choice.\\

\emph{Remark}. The arrow (c) is an involution as the original Legendre transformation was: this symmetry between Lagrange and Hamilton parts is kept! Thus one may construct $\Xi$ from given $\Xi^*$ and describe it locally as a graph of function $\Lambda$ (it will succeed if $\Xi^*$ is not defected). This may be used to construct a Lagrange function for a system with Dirac constrains.\\

The described cases may be visualized by a modified diagram (suppose $U :=V\cap\lbrace f_\alpha(q,v)=0\rbrace $):
\begin{equation}
\xymatrix{
&\Xi \subset \mathbb{P} \left( \mathbb{R} \oplus V \right) \ar@{<->}[r]^{(c)} & \Xi^* \subset \mathbb{P} \left( \mathbb{R} \oplus V^* \right) \ar@{->}[d]^{(d)} & & \\
\Lambda \colon U \to \mathbb{R} \ar@{->}[ur]^{(b)} & f_\alpha(q,v)=0 \ar@{->}[u]^{(b)} & \Sigma \subset  \mathbb{A} \left( V^* \right) \ar@{->}[d]^{(e)} \ar@{->}[dr]^{(e)} &\\
& \ar@{->}[ul]^{(a)} Lagrange\: function\: on\: f_\alpha(q,v)=0 \ar@{->}[u]^{(a)} & Hamilton\: function & Constraints
} \label{Diagram2}
\end{equation}
Note that in this formalism separation of Hamilton function from constraints is artificial: they have the same nature and we extract Hamiltonian due to historical reasons. It is more symmetrical and explains why a linear combination of Hamiltonian and constraints is also a Hamiltonian (see \cite{Dirac}\cite{Tyutin}).
%subsubsection{Defected cases}
%To be defined only by one equation ( like in classic situation by $H = H(p)$ ) $\Xi^*$ must have the codimension 1. However in some situations it would be higher. In this situations $\Xi$ is called {\slshape defected}. Some of this systems with non-convex Lagrange function can be properly described by a new version of Legendre transformation. \\
%Despite the term "defected" provokes to think about this case as special one, a lot of them can be described. Define $\Phi_\Xi \subset \mathbb{P}{V}\times\mathbb{P}^*{V}$ as the submanifold of pairs $(X,P)$, there $X\in\Xi$ and $P=0$ is tangent to $\Xi$ at $X$. For any dimension of $\Xi$, $\Phi_\Xi$ will have dimension $dim(V)-2$, but after projection to $\Xi^*$ it might reduce. Also $\Phi_\Xi = \Phi_{\Xi^*}$, so double dual is the same manifold. Consequently, if $codim(\Xi)>1$, $\Xi^*$ is defected. The image of any point of $\Xi$ is a linear space of dimension $codim(\Xi)-1$. So, the defected manifold must be sliced to linear subspaces.\\
%It is widely known that not all the defected manifolds are cones of a manifold ( in terms of Lagrange systems that means an extra variable, which doesn't enter the Lagrange function ). For example, Segre manifold $\Sigma_{1,2}$ is self-dual and has codimension 2 ( again see \cite{Harris} ).

\subsection{Worldsheet case}

If one wants to work with fields or strings, the 1-dimension time $\mathcal{T}$ must be replaced by a worldsheet with dimension $k>1$.\\ 
Tangent vectors to a path are replaced by decomposable polyvectors, since they are exterior products of tangent vectors. Also any decomposable polyvector can be tangent to some surface. According to this $\Lambda$ is defined for decomposable vectors.\\
Let us denote the subset $\mathcal{D}\subset\Lambda^k T_{t,q}(\mathcal{T \times F} )$ of decomposable polyvectors at point $ (t,q) \in \mathcal{T \times F} $  (it is connected with the Grassmann variety by the formula $\mathbb{P}\mathcal{D}=Gr(dim\mathcal{T},dim\left( \mathcal{T \times F} \right) )$). $\mathcal{D}$ satisfies the same Pl\"ucker relations as the Pl\"ucker embedding of Grassmannian. Let us denote them as $\pi^\alpha = 0 $.\\
In this case (\ref{Diagram2}) is
\begin{equation}
\xymatrix{
&\Xi \subset \mathbb{P} \left( \mathbb{R} \oplus \Lambda^k T_{t,q}(\mathcal{T \times F}) \right) \ar@{<->}[r]^{(c)} & \Xi^* \subset \mathbb{P} \left( \mathbb{R} \oplus \Lambda^k T_{t,q}^*(\mathcal{T \times F}) \right) \ar@{->}[d]^{(d)} &\\
\Lambda \colon \mathcal{D} \to \mathbb{R} \ar@{->}[ur]^{(b)} & \pi^\alpha=0 \ar@{->}[u]^{(b)} & \Sigma \subset  \mathbb{A} \left( \Lambda^k T_{t,q}^*(\mathcal{T \times F}) \right) \ar@{->}[d]^{(e)} \\
& \ar@{->}[ul]^{(a)} Lagrange\: function\: on\: \mathcal{D} \ar@{->}[u]^{(a)} & Hamilton\: function
}
\end{equation}
To obtain a parametric description of $\Xi^*$ let us define a submanifold $\Phi \subset \left( \mathbb{R} \oplus \Lambda^k T_{t,q}(\mathcal{T \times F}) \right) \times \left( \mathbb{R} \oplus \Lambda^k T_{t,q}^*(\mathcal{T \times F}) \right) $ with coordinates $(\Lambda, \xi_I, \Pi, P^I )$ by a system of equations (with N Pl\"ucker relations):
\begin{gather}
X := \Lambda - \Lambda(\xi_I) = 0.  \\
\pi^\alpha(\xi_I) = 0. \\
rank
\begin{pmatrix}
\partial_\Lambda X  & \partial_I X \\
\partial_\Lambda \pi^\alpha  & \partial_I \pi^\alpha \\
\Pi & P^I
\end{pmatrix} \equiv
rank
\begin{pmatrix}
1  & - \partial_I \Lambda \\
0  & \partial_I \pi^\alpha \\
\Pi & P^I
\end{pmatrix} = N + 1. \label{LinearDep}
\end{gather}
The projection of $\Phi $ to $ \mathbb{R} \oplus \Lambda^k T_{t,q}(\mathcal{T \times F}) $ is $\Xi$ and the projection to $ \mathbb{R} \oplus \Lambda^k T_{t,q}^*(\mathcal{T \times F}) $ is $\Xi^*$.\\ 
The equation (\ref{LinearDep}) corresponds to the linear dependence of $(\Pi, P^I)$ from $(\partial_\Lambda X , \partial_I X)$ and $ (\partial_\Lambda \pi^\alpha , \partial_I \pi^\alpha )$\footnote{At a smooth point of $\Xi$ }. The system provides a parametrization of $\Xi^*$ by points of $\Xi$ and coefficients of linear dependency in (\ref{LinearDep}). \\

\emph{Remark} In situation with no Pl\"ucker relations (in cases of scalar field or worldline), (\ref{LinearDep}) contains two rows. Decomposing by the first column one obtains $p_I+\Pi\partial_I \Lambda = 0$. With $\Pi = -1$ that means $P_I = \partial_I \Lambda$.

\section{Examples of usage}
\label{sec:examples}

\subsection{Scalar field}
\label{ssec:SF}

\subsubsection{1+1-dimensional spacetime}
First we should present proper calculation which yields the equation of the surface presented in \ref{ssec:SF11}. The procedure is separated in parts corresponding to the arrows in diagram (\ref{Diagram}). \\
The proposed in section \ref{ssec:SF11} equation of motion $\bigtriangleup \phi = 0$ may be derived for the Lagrange density
\begin{equation}
\mathcal{L} = \frac{1}{2} \left( \partial_0 \phi \partial_0 \phi- \partial_1 \phi \partial_1 \phi \right) .
\end{equation} 
We are going to use the procedure described in section \ref{sec:LT} for $\mathcal{L}$.\\
First let us express Lagrange density in terms of $\mathbb{P}(\mathbb{R}^3)$
coordinates. The tangent polyvector $\upsilon$ to a 2d-surface in a 3d space is
an element of $\Lambda^2(\mathbb{R}^3)$ and is a linear combination of basis
vectors:
\begin{equation}
\upsilon = \overline{X}_{\phi} \dfrac{\partial\:}{\partial x^0} \wedge \dfrac{\partial\:}{\partial x^1} + \overline{X}_1 \dfrac{\partial\:}{\partial x^{\phi}} \wedge \dfrac{\partial\:}{\partial x^0} + \overline{X}_0 \dfrac{\partial\:}{\partial x^{\phi}} \wedge \dfrac{\partial\:}{\partial x^1}, \label{Xdef}
\end{equation} 
so $\overline{X}_0$, $\overline{X}_1$ and $\overline{X}_{\phi}$ are coordinates
on $\Lambda^2(\mathbb{R}^3)$. A tangent polyvector $\upsilon$ to the graph of $\phi(x^0,x^1)$ can be expressed by derivatives of $\phi$:
\begin{equation}
\upsilon = C \left( \dfrac{\partial\:}{\partial x^0} + \partial_0 \phi \dfrac{\partial\:}{\partial \phi} \right) \wedge 
\left( \dfrac{\partial\:}{\partial x^1} + \partial_1 \phi \dfrac{\partial\:}{\partial \phi} \right).
\label{TangPoly}
\end{equation}
Comparing (\ref{Xdef}) and (\ref{TangPoly}), we have
\begin{gather}
\partial_0 \phi = \frac{\overline{X}_0}  {\overline{X}_{\phi} }, \label{PartPhi0} \\
\partial_1 \phi = - \frac{\overline{X}_1}  {\overline{X}_{\phi} }. \label{PartPhi1}
\end{gather}
{\renewcommand{\theenumi}{\alph{enumi}}
\renewcommand{\labelenumi}{(\theenumi)}
\begin{enumerate}
\item
The function $\Lambda$ presented in \ref{sec:LT} is related to $\mathcal{L}$ by
\begin{equation}
\Lambda(\upsilon) = \mathcal{L}(\partial_0 \phi(\upsilon), \partial_1 \phi(\upsilon) ) \cdot dx^0 \wedge dx^1 (\upsilon),
\end{equation}
where $\upsilon$ is a tangent polyvector and $\partial_i \phi(\upsilon)$ are defined by (\ref{PartPhi0}) and (\ref{PartPhi1}).\\
Having 
\begin{equation}
dx^0 \wedge dx^1 (\upsilon) = \overline{X}_{\phi}
\end{equation}
from the definition of $\overline{X}_{\phi}$ in (\ref{Xdef}), we finally obtain
\begin{equation}
\Lambda = \frac{1}{2} \left[ \left( \frac{\overline{X}_0}  {\overline{X}_{\phi} } \right) ^2 - \left( - \frac{\overline{X}_1}  {\overline{X}_{\phi} } \right) ^2 \right] \overline{X}_{\phi} . \label{Lamba}
\end{equation}
\item
The formula (\ref{Lamba}) can be rewritten in terms of zeroes of homogeneous polynomial
\begin{equation}
\overline{X}_{\phi} \Lambda - \frac{1}{2} \overline{X}_0^2 + \frac{1}{2} \overline{X}_1^2 = 0, \label{HomEq}
\end{equation}
so it defines a projective variety $\Xi$ in $\mathbb{P}(\mathbb{R}^4)$ with homogeneous coordinates $[\Lambda : \overline{X}_{\phi} : \overline{X}_0 :\overline{X}_1 ]$.
\item
Now we are going to construct the dual variety for $\Xi$.\\
First we note that $\Xi$ is a quadric. Let us introduce the following notation:
\begin{equation}
X_0 = \Lambda, \quad X_1 = \overline{X}_{\phi}, \quad X_2 = \overline{X}_0, \quad X_3 =  \overline{X}_1
\end{equation}
and
\begin{equation}
G^{\alpha \beta} =
\begin{pmatrix}
0 & \frac{1}{2} & 0 & 0 \\
\frac{1}{2} & 0 & 0 & 0 \\
0 & 0 & - \frac{1}{2} & 0\\
0 & 0 & 0& \frac{1}{2}&
\end{pmatrix}.
\end{equation}
In this notation (\ref{HomEq}) is
\begin{equation}
G^{\alpha \beta} X_\alpha X_\beta =0. \label{Quadric}
\end{equation}
Let us define projective coordinates in the dual projective space as 
$[\Pi : P^{\phi} : P^0 : P^1 ]$ ($\Pi$ is dual to $\Lambda$, $P^{\phi}$ is dual
to $\overline{X}_{\phi}$, etc). Any point of this space defines a hyperplane 
by equation
\begin{equation}
\Pi^\alpha X_\alpha = 0,
\end{equation}
where $\Pi^0 = \Pi$, $\Pi^1 = P^\phi $, $\Pi^2 = P^0 $ and $\Pi^3 = P^1 $. This hyperplane is tangent to $\Xi$ at point $X_\alpha$, iff $X_\alpha$ is a point of $\Xi$ and $\Pi^\alpha$ is (up to a multiplication on nonzero scalar) $\partial_\alpha \left( G^{\beta \gamma} X_\beta X_\gamma \right) = 2 G^{\alpha \beta} X_\beta $. Inverting the last relation (it is a bijection) and substituting $X^\alpha$ to (\ref{Quadric}), we get
\begin{equation}
G_{\alpha \beta} \Pi^\alpha \Pi^\beta = 0.
\end{equation}
This equation can be rewritten in terms of $\Pi $, $P^{\phi}$, $P^0$, $P^1$ as 
\begin{equation}
\Pi P^{\phi} - \frac{1}{2} (P^0)^2 + \frac{1}{2} (P^1)^2 = 0. \label{XiDual}
\end{equation}
This is the equation for $\Xi^*$.
\item
Finally we obtain surface $\Sigma$. It is $\Xi^*$ in the affine map $\Pi = -1$. So its equation is (\ref{XiDual}) with $\Pi = -1$:
\begin{equation}
P^{\phi} + \frac{1}{2} (P^0)^2 - \frac{1}{2} (P^1)^2 = 0. \label{SigmaEq}
\end{equation}
as it was proposed in section \ref{ssec:SF11} ( in notation of that section $P^{\phi} = p_{01} $, $P^{0} = p_{\phi 1} $ and $P^{1} = p_{\phi 0} $ ). So this algorithm leads to equations equivalent to the Klein-Gordon equation obtained in Lagrange formalism.
\item
One of the coordinates can be (at least locally) expressed as explicit function of the others. It can be given by
\begin{equation}
P^{\phi} = -\frac{1}{2} (P^0)^2 + \frac{1}{2} (P^1)^2
\end{equation}
which has the form proposed in (\ref{ssec:SF11}). However this is not essential for the equations of motion. We could stop at (\ref{SigmaEq}).
\end{enumerate}
}

\subsubsection{Arbitrary dimension of spacetime}

 Here we consider the scalar field in n-dimensional spacetime. All the steps have similar ones in  the case of 1+1 spacetime. \\
The Lagrange density is
\begin{equation}
\mathcal{L} = \frac{1}{2} g^{ij} \partial_i \phi \partial_j \phi + \Psi(x,\phi),
\end{equation} 
where $\Psi(x,\phi)$ may contain a mass part and potential energy.\\
To rewrite the action we should obtain the connection between $\partial_i \phi$ and the standard coordinates $\overline{X}_{\phi}$, $\overline{X}_i$ in $ \Lambda^n T( \mathbb{R \times R}^n ) $:
\begin{equation}
\forall \upsilon \in \Lambda^n T( \mathbb{R \times R}^n ) \quad 
\upsilon = \overline{X}_{\phi} \dfrac{\partial\:}{\partial x^0} \wedge ... \wedge \dfrac{\partial\:}{\partial x^n} + \sum\limits_i \overline{X}_i \dfrac{\partial\:}{\partial \phi} \wedge \dfrac{\partial\:}{\partial x^0} ... \wedge \widehat{ \dfrac{\partial\:}{\partial x^i} } \wedge ... \wedge \dfrac{\partial\:}{\partial x^n}.
\end{equation} 
For a tangent polyvector to the graph of $\phi(x^0,...,x^n)$ one can easily obtain the following relations:
\begin{equation}
\partial_i \phi = (-1)^i \frac{\overline{X}_i}  {\overline{X}_{\phi} }. \label{ParPhi}
\end{equation}
After the definition of a new metric
\begin{equation}
\check{g}^{ij} = J_i^k g^{kl} J_j^l, \quad J^i_j = (-1)^i \delta^i_j,
\end{equation}
we obtain
\begin{equation}
\mathcal{L}(\phi, x^0, ..., x^n, \partial_0 \phi(\upsilon), ... , \partial_n \phi(\upsilon) ) = \frac{1}{2} \check{g}^{ij} 
\left( \frac{ \overline{X}_i }{ \overline{X}_{\phi} } \right)
\left( \frac{ \overline{X}_j }{ \overline{X}_{\phi} } \right)
+ \Psi(x,\phi),
\end{equation}
where $\partial_i \phi(\upsilon)$ is defined by the coordinates $\overline{X}_{\phi}$ and $\overline{X}_i$ of polyvector $\upsilon$ tangent to the graph of $\phi(x^0,...,x^n)$ ( by formula \ref{ParPhi}).
{\renewcommand{\theenumi}{\alph{enumi}}
\renewcommand{\labelenumi}{(\theenumi)}
\begin{enumerate}
\item
The function $\Lambda$ is related to $\mathcal{L}$ by
\begin{equation}
\Lambda(\upsilon) = \mathcal{L}(\phi, x^0, ..., x^n, \partial_0 \phi(\upsilon), ... ,\partial_n \phi(\upsilon) ) \cdot dx^0 \wedge ... \wedge dx^n (\upsilon).
\end{equation}
and from the definition of $\overline{X}_{\phi}$
\begin{equation}
dx^0 \wedge ... \wedge dx^n (\upsilon) = \overline{X}_{\phi},
\end{equation}
so
\begin{equation}
\Lambda = \overline{X}_{\phi} \left[ \frac{1}{2} \check{g}^{ij} 
\left( \frac{ \overline{X}_i }{ \overline{X}_{\phi} } \right)
\left( \frac{ \overline{X}_j }{ \overline{X}_{\phi} } \right)
+ \Psi(x,\phi) \right] \label{LambdaN}
\end{equation}
\item
The solutions of equation (\ref{LambdaN}) can be expressed as zeroes of homogeneous polynomial
\begin{equation}
\Lambda \overline{X}_{\phi} - \Psi(x,\phi) \overline{X}_{\phi}^2 - \frac{1}{2} \check{g}^{ij} \overline{X}_{i} \overline{X}_{j} = 0.
\end{equation}
and this is a projective variety $\Xi$ in $\mathbb{P}(\mathbb{R}^{n+2})$.
\item
The next step is to construct the variety dual to $\Xi$.\\
As it was in case of 1+1 scalar field, $\Xi$ is a quadric in projective space $ \mathbb{P} \left( \mathbb{R} \oplus
\Lambda^n T_x( \mathbb{R \times R}^n ) \right) $ with homogeneous coordinates
$[\Lambda: \overline{X}_{\phi}: \overline{X}_{0}: ... : \overline{X}_{n} ]$. The
equation can be rewritten with the help of symmetric bilinear form $G^{\alpha \beta}$:
\begin{equation}
G^{\alpha \beta} X_\alpha X_\beta = 0, 
\quad X_\alpha \in \mathbb{R} \oplus \Lambda^n T_x( \mathbb{R \times R}^n ), \label{LEq} 
\end{equation}
where $ X_0 = \Lambda$, $X_1 = \overline{X}_{\phi}$ and for $ 0 \leqslant i \leqslant n$  $ X_{2+i} = \overline{X}_{i} $. Also $G^{\alpha \beta}$ has the following matrix
\begin{equation}
G^{\alpha \beta} =
\begin{pmatrix}
0 & \frac{1}{2} & 0 & \ldots & 0 \\
\frac{1}{2} & -\Psi(x,\phi) & 0 & \ldots & 0 \\
0 & 0 & & & \\
\vdots & \vdots & &-\frac{1}{2} \check{g}^{ij} & \\
0 & 0 & & &
\end{pmatrix}.
\end{equation}
The tangent hyperspace to that quadric at point $X_\alpha$ is class of equivalence of $
\Pi^\alpha = \partial_\alpha \left( G^{\beta \gamma} X_\beta X_\gamma \right) = 2 G^{\alpha \beta} X_\beta \in \left[ \mathbb{R} \oplus \Lambda^n T_x( \mathbb{R \times R}^n ) \right]^*$. Inverting this relation and using (\ref{LEq}) one
obtains the equation on the dual variety (the union of all tangent hyperplanes):
\begin{equation}
\Pi^2 \Psi(x,\phi) + \Pi P^\phi - \frac{1}{2} \check{g}_{ij} P^i P^j = 0, \label{preHSurf}
\end{equation}
where $\Pi^\alpha = (\Pi,P^\phi,P^0,\:...\:,P^n)$.
\item
Finally we obtain surface $\Sigma$. It is $\Xi^*$ at the affine map $\Pi=-1$.
Making $\Pi$ to take the default value $\Pi = -1$ and labeling the LHS of (\ref{preHSurf}) as $-\eta$ we obtain
\begin{equation}
\eta := P^\phi + \frac{1}{2} \check{g}_{ij} P^i P^j - \Psi(x,\phi) = 0. \label{HSurfDef}
\end{equation}
And we have $\Sigma:\; \eta =0$.
\item
We don't need an explicit dependence of one coordinate from the others, so we omit this step.
\end{enumerate}
}

Further we derive the equations of motion yielded by the surface $\eta = 0$.\\
To eliminate $\omega\vert_{\eta=0}$ the polyvector with a nondegenerate projection on the spacetime 
\begin{equation}
\Xi = C \cdot \bigwedge\limits_i \left(
\dfrac{\partial\:}{\partial x^i} +
f_i \dfrac{\partial\:}{\partial \phi} + 
\pi_i^k \dfrac{\partial\:}{\partial P^k} +
\pi_i^\phi \dfrac{\partial\:}{\partial P^\phi}
\right)
\end{equation}
must satisfy relation
\begin{equation}
\exists \alpha \quad i_\Xi \omega = \alpha d\eta
\Longleftrightarrow
i_{\Xi \wedge v} \omega = \alpha L_v \eta.
\label{HSurf}
\end{equation}
For the basis vectors in (\ref{HSurf}) we have
\begin{align}
\alpha & \equiv & \alpha \dfrac{\partial\eta}{\partial P^\phi} & = & (-1)^{n+1}, \label{HE1}\\
\alpha \check{g}_{ij} P^j & \equiv & \alpha \dfrac{\partial\eta}{\partial P^i} & = & (-1)^{n+1} \cdot (-1)^i f_i, \label{HE2}\\
- \alpha \dfrac{\partial\Psi}{\partial \phi} & \equiv & \alpha \dfrac{\partial\eta}{\partial \phi} & = & (-1)^n \sum\limits_k (-1)^k \pi^k_k, \label{HE3} \\
- \alpha \dfrac{\partial\Psi}{\partial x^k} & \equiv & \alpha \dfrac{\partial\eta}{\partial x^k} & = & (-1)^n \left[ \pi^{\phi}_k + \sum\limits_i (-1)^i [ \pi^i_k f_i -\pi^i_i f_k ] \right]. \label{HE4}
\end{align}
The form of $\Xi$ means that $f_i = \dfrac{\partial\phi}{\partial x^i}$ and $\pi^k_i = \dfrac{\partial P^k}{\partial x^i}$. So (\ref{HE2}) and (\ref{HE1}) lead to
\begin{equation}
P^i = \sum\limits_j \check{g}^{ij} (-1)^j \dfrac{\partial\phi}{\partial x^j} = (-1)^i g^{ij} \dfrac{\partial\phi}{\partial x^j}.
\end{equation}
Consequently, (\ref{HE3}) means
\begin{equation}
g^{ij} \dfrac{\partial^2\phi}{\partial x^i \partial x^j} = \dfrac{\partial\Psi}{\partial \phi}
\Longleftrightarrow
\bigtriangleup \phi = \dfrac{\partial\Psi}{\partial \phi}. \label{Res}
\end{equation}
This is the only differential equation on $\phi$: as $\eta \equiv 0$, $P^\phi \equiv -\frac{1}{2} \check{g}_{ij} P^i P^j + \Psi(x,\phi)$ and this makes (\ref{HE4}) equivalent to (\ref{HE3}):
$$
\pi^\phi_k - \dfrac{\partial\Psi}{\partial x^k} + \sum\limits_i (-1)^i [ \pi^i_k f_i - \pi^i_i f_k ] = 0
\Longleftrightarrow
$$
$$
-\check{g}_{ij} P^i \dfrac{\partial P^j}{\partial x^k} + \dfrac{\partial\Psi}{\partial x^k} + \dfrac{\partial\Psi}{\partial \phi} \dfrac{\partial\phi}{\partial x^k} - \dfrac{\partial\Psi}{\partial x^k} + \sum\limits_i (-1)^i [\dfrac{\partial P^i}{\partial x^k} (-1)^i \check{g}_{ij}P^j - \pi^i_i \dfrac{\partial\phi}{\partial x^k} ] = 0
\Longleftrightarrow
$$
$$
\dfrac{\partial\phi}{\partial x^k} \left[ \dfrac{\partial\Psi}{\partial \phi} - \sum\limits_i (-1)^i \pi^i_i \right] = 0.
$$
For a field with mass $m$ we have $ \Psi(x,\phi) \equiv - \frac{1}{2}m^2 \phi^2 $, so
\begin{equation}
\bigtriangleup \phi = -m^2\phi.
\end{equation}

\subsection{Electrodynamics in 1+1}
\label{ssec:ED}

In this section an analogous procedure will be done for Maxwell-like action in 1+1 dimension spacetime. The new difficulty will be observed: we will have nontrivial Pl\"ucker relations. \\

\emph{Notation.} In this section underlined indices ($ \underline{i}$, $\underline{0}$, $\underline{1} $, etc) are used for the coordinates of fields (they replace standard coordinates in mechanics) and non-underlined represent coordinates of the parameterizing manifold (space-time here).\\
In skew forms following alphabetical order is used: underlined indices are placed before non-underlined. Thus, the natural $(k+1)$-form on $\Lambda^k T(\mathcal{F\times T})$ is written as 
\begin{equation} 
\omega = \sum\limits_{ 
\begin{split} 
\underline{i_1}<...<\underline{i_r},\\ j_1<...<j_{k-r}. \end{split} }
dp_{ \underline{i_1 ... i_r} j_1 ... j_{k-r} } \wedge d\phi^{\underline{i_1}} \wedge ... \wedge d\phi^{\underline{i_r}} \wedge dx^{j_1} \wedge ... \wedge dx^{j_{k-r}},
\end{equation} 
where $\phi^{\underline i}$ are field coordinates, $ x^j $ are spacetime coordinates.\\
The Lagrange function is given
\begin{equation}
\mathcal{L} = C_0 F_{\mu \nu} F^{\mu \nu} + \Phi(A_{\mu}, x^{\mu}).
\end{equation}
Further we usually write $ \phi^{\underline{i} }$ instead of $ A_i $.\\
Again we investigate the relation between $\partial_i \phi^{\underline{j} }$ and $X^{\underline{01}}, X^{\underline{0}0}, X^{\underline{0}1}, X^{\underline{1}0}, X^{\underline{1}1}, X^{01}$, the coordinates in $\Lambda^2 T(\mathbb{R}^2 \times \mathbb{R}^2)$. The following relations are satisfied:
\begin{gather}
\frac{X^{\underline{01}}}{X^{01}} = \dfrac{\partial\phi^{\underline{0} }}{\partial x^0} \dfrac{\partial\phi^{\underline{1} }}{\partial x^1} - \dfrac{\partial\phi^{\underline{0} }}{\partial x^1} \dfrac{\partial\phi^{\underline{1} }}{\partial x^0} , \\
\frac{X^{\underline{0}0}}{X^{01}} = -\dfrac{\partial\phi^{\underline{0} }}{\partial x^1}, \label{EM1}\\
\frac{X^{\underline{0}1}}{X^{01}} = \dfrac{\partial\phi^{\underline{0} }}{\partial x^0}, \\
\frac{X^{\underline{1}0}}{X^{01}} = -\dfrac{\partial\phi^{\underline{1} }}{\partial x^1}, \\
\frac{X^{\underline{1}1}}{X^{01}} = \dfrac{\partial\phi^{\underline{1} }}{\partial x^0}. \label{EM2} \\
\left( \frac{X^{\underline{i}j}}{X^{01}} = \varepsilon^{kj} \dfrac{\partial\phi^{\underline{i} }}{\partial x^k} \right) \notag
\end{gather}
Using (\ref{EM1})-(\ref{EM2}) $\mathcal{L}$ can be written
\begin{equation}
\mathcal{L} = C\left(\frac{X^{\underline{0}1} + X^{\underline{1}0}}{X^{01}} \right)^2 + \Phi, \quad C = 2C_0.
\end{equation}
\begin{equation}
\Lambda = X^{01}\left[ C\left( \frac{X^{\underline{0}1} + X^{\underline{1}0}}{X^{01}} \right)^2 + \Phi \right]
\Longleftrightarrow
\Lambda X^{01} - C (X^{\underline{0}1} + X^{\underline{1}0})^2 - \Phi [X^{01}]^2 = 0. \label{LEM}
\end{equation}
The LHS of (\ref{LEM}) we denote by $F$.\\
In contrast with the example in the first section not all the vectors in $\Lambda^2 T(\mathbb{R}^2 \times \mathbb{R}^2)$ can be tangent bivector to the surface, but only decomposable ones. The conditions equivalent to decomposability are called Pl\"ucker relations. In this case there is only one of them:
\begin{equation}
X^{\underline{01}} X^{01} - X^{\underline{0}0} X^{\underline{1}1} + X^{\underline{0}1} X^{\underline{1}0} = 0. \label{PEM}
\end{equation}
The LHS of (\ref{PEM}) we denote by $\pi$.\\
Equations $F = 0, \; \pi = 0 $ define the variety in $\mathbb{P} \left( \Lambda^2 T_x(\mathbb{R}^2 \times \mathbb{R}^2) \right)$. The element of the dual variety representing the tangent hyperspace at $X$ is the class equivalence of covector $P = \alpha dF\vert_X + \beta d\pi\vert_X$. After excluding all variables except for the coordinates of $P$, the following equation for the dual variety is derived
\begin{equation}
(4\Pi C + P_{\underline{01}} )
\left[ P_{\underline{01}} ( \Pi \Phi +P_{01} ) - P_{\underline{0}0} P_{\underline{1}1} + P_{\underline{0}1} P_{\underline{1}0} \right] + 
\Pi C \left( P_{\underline{0}0} + P_{\underline{1}1} \right)^2= 0. \label{HSurfEM}
\end{equation}
One can see (\ref{HSurfEM}) as deformed Pl\"ucker relation: with $C = 0, \; \Pi = 0$ (it means $\mathcal{L} \equiv 0$) it transforms to an analogue of (\ref{PEM}).\\
The LHS of (\ref{HSurfEM}) is denoted by $\eta$ and plays the same role as in the case of (\ref{HSurfDef}). The equations on a polyvector with nondegenerative projection on spacetime 
\begin{equation}
\begin{split}
\Xi = C \cdot \left( \dfrac{\partial\:}{\partial x^0} +
f_0^{\underline{i} } \dfrac{\partial\:}{\partial\phi^{\underline{i} }} + 
\pi_{{\underline{01} };0} \dfrac{\partial\:}{\partial P_{\underline{01} }} +
\pi_{{\underline{i} }j;0} \dfrac{\partial\:}{\partial P_{{\underline{i} }j}} +
\pi_{01;0} \dfrac{\partial\:}{\partial P_{01}} \right) 
\wedge \\ \wedge
\left(\dfrac{\partial\:}{\partial x^1} +
f_1^{\underline{i} } \dfrac{\partial\:}{\partial\phi^{\underline{i} }} + 
\pi_{{\underline{01} };1} \dfrac{\partial\:}{\partial P_{\underline{01} }} +
\pi_{{\underline{i} }j;1} \dfrac{\partial\:}{\partial P_{{\underline{i} }j}} +
\pi_{01;1} \dfrac{\partial\:}{\partial P_{01}} \right) 
\end{split}
\end{equation}
with basis vectors $\dfrac{\partial\:}{\partial P_{{\underline{i} }j}}$ and $\dfrac{\partial\:}{\partial P_{01}}$ are ($\Pi$ is considered to be constant):
\begin{align}
\alpha P_{\underline{01}} (4 \Pi C + P_{\underline{01}} ) & \equiv & \alpha \dfrac{\partial\eta}{\partial P_{01}} & = & 1, \label{HEEM1}\\
\alpha \left[ - P_{\underline{1}1} (4 \Pi C + P_{\underline{01}} ) + 2\Pi C \left( P_{\underline{0}0} + P_{\underline{1}1} \right) \right] & \equiv & \alpha \dfrac{\partial\eta}{\partial P_{\underline{0}0}} & = & -f^{\underline{0} }_1 , \label{HEEM2}\\
\alpha P_{\underline{1}0} (4 \Pi C + P_{\underline{01}} ) & \equiv & \alpha \dfrac{\partial\eta}{\partial P_{\underline{0}1}} & = & f^{\underline{0} }_0, \label{HEEM3}\\
\alpha P_{\underline{0}1} (4 \Pi C + P_{\underline{01}} ) & \equiv & \alpha \dfrac{\partial\eta}{\partial P_{\underline{1}0}} & = & -f^{\underline{1} }_1, \label{HEEM4}\\
\alpha \left[ - P_{\underline{0}0} (4 \Pi C + P_{\underline{01}} ) + 2\Pi C \left( P_{\underline{0}0} + P_{\underline{1}1} \right) \right] & \equiv & \alpha \dfrac{\partial\eta}{\partial P_{\underline{1}1}} & = & f^{\underline{1} }_0, \label{HEEM5}\\
\alpha \left[ \left( P_{\underline{01}} ( \Pi \Phi +P_{01} ) - P_{\underline{0}0} P_{\underline{1}1} + P_{\underline{0}1} P_{\underline{1}0} \right) + (4 \Pi C + P_{\underline{01}} )( \Pi \Phi +P_{01} ) \right] & \equiv &  \alpha \dfrac{\partial\eta}{\partial P_{\underline{01}}} & = & f^{\underline{0} }_0 f^{\underline{1} }_1 - f^{\underline{0} }_1 f^{\underline{1} }_0. \label{HEEM6}
\end{align}

\emph{Remark}. Extracting $f^{\underline{i} }_j$ from (\ref{HEEM1})-(\ref{HEEM5}) and using $\eta \equiv 0$ we get
\begin{equation}
\alpha \left[ \left( P_{\underline{01}} ( \Pi \Phi +P_{01} ) - P_{\underline{0}0} P_{\underline{1}1} + P_{\underline{0}1} P_{\underline{1}0} \right) + (4 \Pi C + P_{\underline{01}} )( \Pi \Phi +P_{01} ) \right] = f^{\underline{0} }_0 f^{\underline{1} }_1 - f^{\underline{0} }_1 f^{\underline{1} }_0 -  2  \alpha \Pi \Phi ( P_{\underline{01}} + 2 \Pi C ).
\end{equation}
Thus the equation (\ref{HEEM6}) is a consequence of (\ref{HEEM1})-(\ref{HEEM5}) and (\ref{HSurfEM}) iff $\Phi = 0$ or $P_{\underline{01}} + 2 \Pi C = 0 $.\\
For basis vectors $\dfrac{\partial\:}{\partial\phi^{\underline{i} }}$:
\begin{align}
\alpha \Pi \dfrac{\partial\Phi}{\partial \phi^{\underline{0} }} P_{\underline{01}} (4 \Pi C + P_{\underline{01}} ) & \equiv & \alpha \dfrac{\partial\eta}{\partial \phi^{\underline{0} }} & = & -\pi_{{\underline{01} };0} f^{\underline{1} }_1 + \pi_{{\underline{01} };1} f^{\underline{1} }_0 + \pi_{{\underline{0} }0;1} - \pi_{{\underline{0} }1;0}, \label{HEEM7} \\
\alpha \Pi \dfrac{\partial\Phi}{\partial \phi^{\underline{1} }} P_{\underline{01}} (4 \Pi C + P_{\underline{01}} ) & \equiv & \alpha \dfrac{\partial\eta}{\partial \phi^{\underline{1} }} & = & \pi_{{\underline{01} };0} f^{\underline{0} }_1 - \pi_{{\underline{01} };1} f^{\underline{0} }_0 + \pi_{{\underline{1} }0;1} - \pi_{{\underline{1} }1;0} .
\label{HEEM8}
\end{align}
Also for $\dfrac{\partial\:}{\partial x^i } $:
\begin{align}
\alpha \Pi \dfrac{\partial\Phi}{\partial x^0} P_{\underline{01}} (4 \Pi C + P_{\underline{01}} ) & \equiv & \alpha \dfrac{\partial\eta}{\partial x^0} & = & \pi_{{\underline{0} 0};0} f^{\underline{0} }_1 - \pi_{{\underline{0} 0};1} f^{\underline{0} }_0 +\pi_{{\underline{1} 0};0} f^{\underline{1} }_1 - \pi_{{\underline{1} 0};1} f^{\underline{1} }_0 - \pi_{01;0}, \label{HEEM9}  \\
\alpha \Pi \dfrac{\partial\Phi}{\partial x^1} P_{\underline{01}} (4 \Pi C + P_{\underline{01}} ) & \equiv & \alpha \dfrac{\partial\eta}{\partial x^1} & = & \pi_{{\underline{0} 1};0} f^{\underline{0} }_1 - \pi_{{\underline{0} 1};1} f^{\underline{0} }_0 +\pi_{{\underline{1} 1};0} f^{\underline{1} }_1 - \pi_{{\underline{1} 1};1} f^{\underline{1} }_0 - \pi_{01;1}. \label{HEEM10}
\end{align}
Equations (\ref{HEEM1})-(\ref{HEEM6}) and (\ref{HEEM7})-(\ref{HEEM10}) are the Hamilton equations for this field.\\
Using
\begin{equation}
\dfrac{\partial\phi^{\underline{i} }}{\partial x^j} = f^{\underline{i} }_j; \;
\dfrac{\partial P_{01} }{\partial x^i} = \pi_{01;i}; \;
\dfrac{\partial P_{\underline{i} j} }{\partial x^k} = \pi_{\underline{i} j;k}; \;
\dfrac{\partial P_{\underline{01}} }{\partial x^k} = \pi_{\underline{01};k},\label{Deriv}
\end{equation}
and (\ref{HEEM1})-(\ref{HEEM5}) the equations (\ref{HEEM7}),(\ref{HEEM8}) can be transformed to
\begin{align}
\Pi \dfrac{\partial\Phi}{\partial \phi^{\underline{0} }} & = 2 \Pi C \dfrac{\partial\:}{\partial x^1 } \left[ \dfrac{\partial\phi^{\underline{0} }}{\partial x^1 } - \dfrac{\partial\phi^{\underline{1} }}{\partial x^0 } \right], \label{Res1} \\
\Pi \dfrac{\partial\Phi}{\partial \phi^{\underline{1} }} & = -2 \Pi C \dfrac{\partial\:}{\partial x^0 } \left[ \dfrac{\partial\phi^{\underline{0} }}{\partial x^1 } - \dfrac{\partial\phi^{\underline{1} }}{\partial x^0 } \right]. \label{Res2}
\end{align}
Defining RHS of (\ref{HEEM7}), (\ref{HEEM8}), (\ref{HEEM9}) and (\ref{HEEM10}) as $\Xi_1$, $\Xi_2$, $\Xi_3$ and $\Xi_4$ respectively, the following relations can be written from (\ref{HEEM1})-(\ref{HEEM5}), (\ref{Deriv}) and $\eta \equiv 0$:
\begin{align}
\Xi_3 + \dfrac{\partial\phi^{\underline{1} }}{\partial x^0} \Xi_2 + \dfrac{\partial\phi^{\underline{0} }}{\partial x^0} \Xi_1& \equiv \Pi \left[ \dfrac{\partial\Phi}{\partial x^0} + \dfrac{\partial\phi^{\underline{0} }}{\partial x^0} \dfrac{\partial\Phi}{\partial \phi^{\underline{0} }} + \dfrac{\partial\phi^{\underline{1} }}{\partial x^0} \dfrac{\partial\Phi}{\partial \phi^{\underline{1} }} \right] ,  \\
\Xi_4 + \dfrac{\partial\phi^{\underline{1} }}{\partial x^1} \Xi_2 + \dfrac{\partial\phi^{\underline{0} }}{\partial x^1} \Xi_1& \equiv \Pi \left[ \dfrac{\partial\Phi}{\partial x^1} + \dfrac{\partial\phi^{\underline{0} }}{\partial x^1} \dfrac{\partial\Phi}{\partial \phi^{\underline{0} }} + \dfrac{\partial\phi^{\underline{1} }}{\partial x^1} \dfrac{\partial\Phi}{\partial \phi^{\underline{1} }} \right].
\end{align}
So (\ref{HEEM9}) and (\ref{HEEM10}) are satisfied if (\ref{HEEM7}) and (\ref{HEEM8}) hold.\\
Finally we obtain that equations on $\phi $ are (\ref{Res1}) and (\ref{Res2}). All other equations either define $P_{01}$, $P_{\underline{i} j}$ and $P_{\underline{01}}$ as functions of $\dfrac{\partial\phi^{\underline{i} }}{\partial x^j}$ or are equivalent to these two.\\
For electromagnetism without currents $\Phi \equiv 0$, thus (\ref{Res1}) and (\ref{Res2}) lead to
\begin{equation}
\dfrac{\partial\phi^{\underline{0} }}{\partial x^1 } - \dfrac{\partial\phi^{\underline{1} }}{\partial x^0 } = const.
\end{equation}

\section{Conclusion}

In the present paper we introduced a new modification of the Hamiltonian formalism. The first feature is that it may be obtained for systems with a worldsheet in place of the worldline of conventional formalism. The equations of motion are expressed in terms of differential form (an analogue of symplectic form) degeneration. That simple type of equation has a benefit of expressing the action (at least locally) as integral of a differential form. The action naturally arises in this form on any subsurface $\Sigma$ of the modified phase space $\Lambda^{dim\mathcal{T}} T^*\left( \mathcal{T \times F} \right)$ and all the dynamics are defined by $\Sigma$. Also we constructed the method to obtain for any Lagrange function $L$ such a subsurface $\Sigma$ that it provides the same dynamics as the Euler-Lagrange equations for $L$.\\
The feature of this method is that it works for Lagrange functions defined only on a subset of velocities, defined by homogeneous constraints $f_\alpha (q,v) = 0$. This allowed to apply this method to Lagrange functions in case of field theory despite Lagrange function being defined only for decomposable polyvectors.\\
In our description we naturally obtained a formalism for degenerate systems. They may be considered as the dual case for Lagrange function defined on a subset: in this cases a variety with a codimention higher than 1 appears on different sides of provided involution.\\
However it is unclear if the method may be modified to describe Lagrange systems with higher derivatives. These $L$'s may be considered as functions on subbundle of tangent bundle of $T^k\mathcal{M}$: for second derivatives it contains all the elements $(x,v,\dot{x},\dot{v})$ with equation $\dot{x} = v$. Thus the functions are not homogeneous on the tangent space. \\
Also we don't know if the equations of motion can be expressed in terms Nambu brackets (see \cite{Takhtajan}, \cite{CurtZach}). It looks likely to have a link with Nambu mechanics because of the appearence of the $n$-form.

\section*{Acknowledgements}

The author is grateful to Alexei Morozov, Alexei Sleptsov, Petr Dunin-Barkowski and Alexandra Anokhina for discussions and helpful remarks.\\
Our work is partly supported by Ministry of Education and Science of Russian Federation under contract 8207, by NSh-3349.2012.2 and grant RFBR 13-02-00478.

\end{document}